\documentclass[%
 reprint,
superscriptaddress,
 amsmath,amssymb,
 aps,
]{revtex4-1}

\usepackage{graphicx}
\usepackage{dcolumn}
\usepackage{bm}
\usepackage{siunitx}
\usepackage{color}

\begin{document}

\preprint{APS/123-QED}


\title{Turning drops into bubbles: Elastic cavitation by diffusion}

\author{M.A.Bruning}
\email{m.a.bruning@utwente.nl}
\affiliation{%
 Physics of Fluids Group, Faculty of Science and Technology, Mesa+ Institute, University of Twente, 7500 AE Enschede, The Netherlands
}%
 \author{M. Costalonga}
\affiliation{Department of Mechanical Engineering, Massachusetts Institute of Technology, Cambridge, Massachusetts, USA}
\author{J. H. Snoeijer}%
\author{A. Marin}
 
\affiliation{%
 Physics of Fluids Group, Faculty of Science and Technology, Mesa+ Institute, University of Twente, 7500 AE Enschede, The Netherlands
}%

\date{\today}

\begin{abstract}

Some members of the vegetal kingdom can achieve surprisingly fast movements making use of a clever combination of evaporation, elasticity and cavitation. In this process, enthalpic energy is transformed into elastic energy and suddenly released in a cavitation event which produces kinetic energy. Here we study this uncommon energy transformation by a model system: a droplet in an elastic medium shrinks slowly by diffusion and eventually transforms into a bubble by a rapid cavitation event. The experiments reveal the cavity dynamics over the extremely disparate timescales of the process, spanning 9 orders of magnitude. We model the initial shrinkage as a classical diffusive process, while the sudden bubble growth and oscillations are described using an inertial-(visco)elastic model, in excellent agreement with the experiments. Such a model system could serve as a new paradigm for motile synthetic materials.
%

\end{abstract}

\maketitle



According to the kids britannica encyclopedia: ``Living things have the ability to move in some way without outside help. The movement may consist of the flow of material within the organism or external movement of the organism or parts of the organism.''\footnote{https://kids.britannica.com/students/article/living-things/275509/225526-toc}
Unlike animal cells, the swelling and shrinkage of plant cells are fundamental for the motion of the whole body of the plant. The key is the wall surrounding the cells: a thin but stiff wall that allows the cells to substain large pressure differences \cite{Dumais2012}.
Motion in plants often occurs over extremely separated time scales, one in the range of hours and/or days, related with tissue swelling/shrinkage and the other in the order of fractions of milliseconds, related with mechanical or thermodynamics instabilities. This fast motion typically involves the storage of elastic energy in the system and its quick (and often dramatic) release. One example of such quick release of energy are those triggered by elastic instabilities, like the snapping of the Venus flytrap \cite{Forterre2005}. It can also involve more violent phenomena as cavitation, as in the fern sporangia \cite{Noblin2012}: the fern's cells shrink when they dehydrate, deforming the whole leave and accumulating elastic energy. This energy is then quickly released in a sudden cavitation event inside several of the fern's cells, which restores the elastic energy stored in the fern and catapults the spores at large distances \cite{Noblin2012}. 
A synthetic analogue system was studied by Vincent et al. \cite{Vincent2012,Vincent2014} using laser-induced cavitation in hydrogels. However, owing to the extreme separation of timescales and the relatively small deformations in those experiments, many aspects regarding the elastic cavitation dynamics remain unknown.
 

\begin{figure}[b!]
  \includegraphics[width=0.48\textwidth]{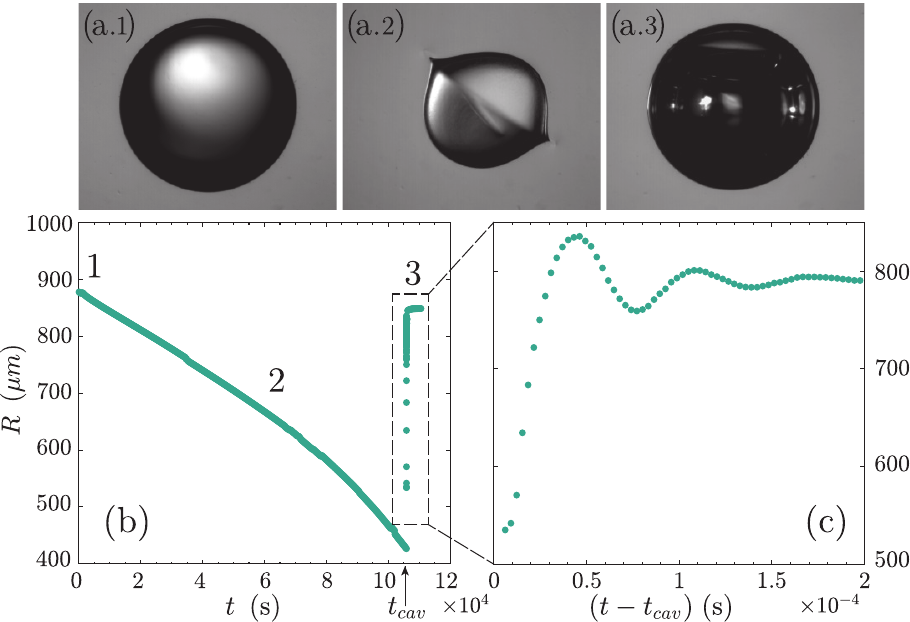}
\vspace{-3mm} \caption{ Different phases in the life of a droplet in a dry and elastic medium: (a.1) The initial water drop in a PDMS-based gel starts to shrink isotropically by evaporation. (a.2) While shrinking, the elastic medium is deformed, certain amount of elastic energy is released by creasing at the solid-liquid interface. (a.3) Eventually, enough negative elastic pressure is accumulated and a cavitation event occurs: a bubble of approximately same size as the original liquid droplet is created. (b) The curve shows the evolution of the effective droplet radius in time and illustrates the disparity of time scales, from a day to a fraction of a millisecond. The corresponding instants for the images in the (a)-panels are also indicated. (c) A zoom of the bubble radius dynamics just after cavitation occurred at $t_{cav}$. \label{fig:1} }
\end{figure}

In this Letter, we unveil the influence of the surrounding medium's viscoelasticity on the dynamics of a fern's cell analogue. Our simplified ``syntethic plant cell'' \cite{Wheeler2008, Vincent2012} consists of a droplet in an elastic medium shrinking due to a slow diffusive process, followed by a rapid cavitation. This system allows us to monitor the whole dynamics of such complex process, from the day-long fading of the liquid droplet, to the millisecond birth and growth of the cavitation bubble. Moreover, we show that these features are accurately captured by a modified Rayleigh-Plesset equation, accounting for the viscoelastic properties of the medium.



The system under study is a single millimetric water droplet trapped in an elastic medium. The gel used is Dow Corning PDMS Sylgard184 mixed in a 1:10 ratio (curing agent:base polymer). The static shear modulus of this gel is 0.7 MPa, as measured using a rheometer (Anton Paar MCR 502). Figure \ref{fig:1}a.1 shows an image of the droplet in its initial state. A transparent containing box is filled with uncured PDMS and placed on top of a heating element (70$^\circ$C). After an initial pre-curing period of the gel of 15 minutes a water droplet is inserted in the center of the box, either using a micro-pipette or a tapered capillary, depending on the desired droplet size $R_0$. The injection of the droplet at this moment allows us to locate the droplet steadily at a given location, while the polymers are still not fully reticulated. This experimental setup has been specifically designed to capture both the whole droplet shrinkage process {(spanning approximately a day)} and the final cavitation event {(which occurs in milliseconds)} in one single experiment with good temporal and spatial resolution. To this end, we use two side view cameras, one recording at one frame per minute (Ximea MQ013MG-ON) and a high speed camera (Photron SAX-2), mounted in perpendicular direction. The high speed camera is installed to capture the dynamics of the cavitation event, at the final stage of the experiment. Image triggering is used to start recording at framerates in a range from 360.000 to 450.000 fps. The duration of the full experiment varies from 3 hours to 50 hours depending on the initial drop radius $R_0$ (from 200 to \SI{1000}{\micro\meter}).

A typical experiment exhibits three stages, as shown in Fig. \ref{fig:1}a. At first, the droplet shrinks isotropically due to diffusion of water vapour into the surrounding PDMS (Fig. \ref{fig:1}a.1). 
Then, at a certain time the droplet looses its spherical shape due to an elastic instability: creasing ~\cite{Chen2014, Cai2010, Weiss2013, Jin2015} (Fig. \ref{fig:1}a.2). During evaporation of the water droplet, the droplet size decreases and therefore stress builds up at the interface with the elastic material, which is partly released by the formation of surface folds. In a somewhat different setup, Milner et al. \cite{Milner2017} studied the onset of creasing for droplets immersed in gels. In the present work, creases are observed below a critical radius $R/R_0$ = 0.73$\pm$0.05, in agreement with ~\cite{Milner2017}.
Finally, as the droplet further evaporates, a negative pressure builds up inside the drop due to the tensile stresses exerted on the cavity, ultimately leading to cavitation (Fig. \ref{fig:1}a.3).
The final radius of the bubble is close to the initial droplet radius (0.98 $R_0$), such that plastic deformation is negligible. An example of how the radius $R(t)$ typically evolves over time is provided in Fig. \ref{fig:1}b. After the creasing, the droplet is no longer spherical and we characterize its size with an effective radius using its projected area. The shrinking process is not altered substantially by the breaking of spherical symmetry, illustrated by the smoothness of $R(t)$ in Figure \ref{fig:1}b. Finally, the cavitation event is highlighted in the zoom in Fig. \ref{fig:1}c. A key feature of the experiment is the disparity of time scales in this single experiment, which span 9 orders of magnitude. 

\begin{figure}
  \includegraphics[width=0.48\textwidth]{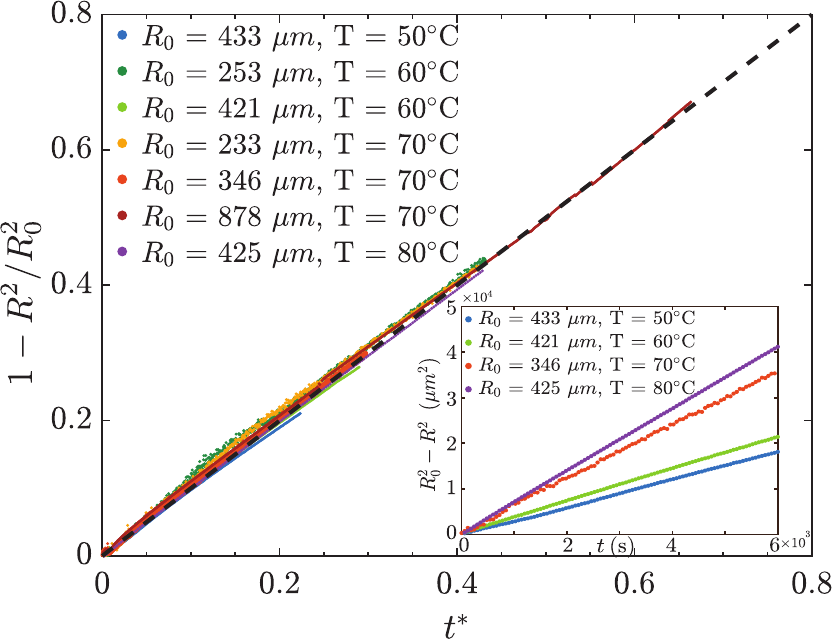}
\vspace{-3mm} \caption{Evaporation of a water droplet in PDMS. The inset shows the data for evaporation at different temperatures, the slope indicates the evaporation rate. The main figure shows the dimensionless $1-R^2/R_0^2$ as a function of $t^* = t/t_F$ for different initial radii and temperatures.   \label{fig:  evap} }
\end{figure}

We now proceed with a more detailed analysis of each of the stages, starting with the slow droplet shrinkage. We hypothesise that this is a diffusion-limited evaporation process in thermal equilibrium, since the liquids are immiscible and no convection is present in the system. Rewriting the droplet radius $R$ as a function of time $t$, this predicts ${R_0}^2-R^2=2D\Delta c\;t/\rho $~\cite{Gelderblom2011}. Here $R_0$ represents the initial droplet size, $\rho$ the density of water in its liquid phase, $D$ the diffusivity of the water vapour in the gel and $\Delta c=c^*-c_\infty$ the concentration gradient between the interface and the far field (the last two properties are temperature dependent). The inset in Fig. \ref{fig:  evap} shows the typical change in radius $R_0^2-R^2$ versus time for different experimental conditions. The linear trend over a wide range of temperatures and ambient humidities (not shown here) confirms the diffusive behaviour.
To quantitatively test the diffusion model, we represent the data in dimensionless form by scaling the radius as $R/R_0$ and time as $t^* = t/t_F = t\;2D\Delta c/(\rho_wR_0^2)$. For the latter scaling, we take the diffusivity $D$ at different temperatures from \cite{Harley2012}, while $\Delta c$ is obtained by fitting the slopes in the inset in Fig. \ref{fig:  evap}. Following this approach we achieve a collapse of the data, shown in the main panel, with values $\Delta c$ = 40-50 mol/m$^3$. These are in good agreement with data from \cite{Randall2005}. Hence, we conclude that the shrinkage of the droplet inside this elastic material is governed by a relatively simple diffusion-limited process.

%
%

 \begin{figure}
  \includegraphics[width=0.48\textwidth]{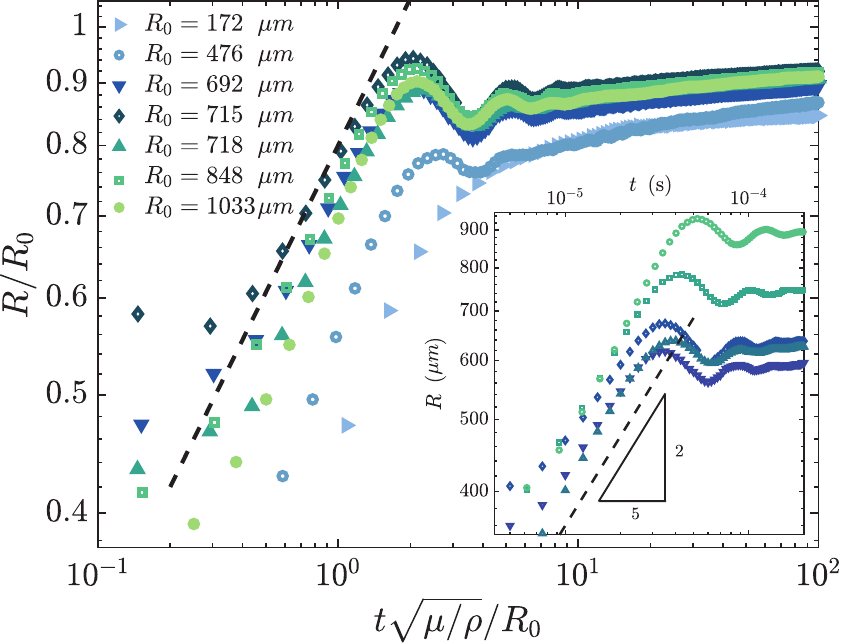}
\vspace{-3mm} \caption{Dynamics of the cavitation bubble. The inset shows the growth of the cavitation for bubble for different initial droplet radii. The main figure shows the non-dimensionalized growth dynamics $R/R_0$ as a function of $t\sqrt{\mu /\rho}/R_0$ (with $\mu=1.7 MPa$, see text). In both figures a power law of 2/5 is displayed. The same legend applies for the two figures. \label{fig:  dyna} }
\end{figure}

The evaporation of the water droplet eventually leads to a very rapid cavitation event, the final stage in the droplet's life, at typical values of $R/R_0=0.48\pm 0.1$. The cavitation occurs due to the tensile elastic stress that builds up in the gel. To estimate this stress, we consider the elastic energy for a spherical cavity, which can be written as $\epsilon_{el} = 4\pi \mu R_0^3 f(\xi) $. In this expression $\mu$ is the static shear modulus, $\xi = R/R_{0}$ the scaled bubble radius, and for a neo-Hookean solid the dimensionless function is $f(\xi)=\frac{5}{6}\xi^3-\xi^2-\frac{1}{3}+\frac{1}{2\xi}$~\cite{Gent1991, Gent1959}. The minimal energy appears at $\xi = 1$, at which the elastic medium is undeformed with respect to its reference state, and a tensile elastic stress appears for $\xi <1 $. The (negative) pressure inside the cavity can then be computed using the virtual work principle $\Delta p \delta V_{cav} = \delta \epsilon_{el} + \gamma \delta A_{cav}$, where we also introduced surface tension $\gamma$, and the volume $V_{cav}$ and area $A_{cav}$ of the cavity/drop. Using the spherical approximation for the cavity, estimating the virtual work from $\delta R$, one obtains the pressure inside the cavity

\begin{equation}
p_{cav} = p_\infty+\frac{2\gamma}{R}+\frac{\mu}{{\xi}^2}f'(\xi), 
\label{p_bubble}
\end{equation}
where $p_\infty$ is the atmospheric pressure far away from the droplet. 
Upon cavitation, $\xi$ reaches typical values of $\xi$ = 0.48, which results in $p_{cav}$ close to -8 MPa.
This pressure is an order of magnitude larger than the static shear modulus, owing to the nonlinearities associated to large deformation. However, given that some of the stress is released by the formation of creases, the calculation above should be seen as an upper estimate for the cavitation pressure. Interestingly, these values are in the same range as measured for the Fern Sporangium, in which cavitation occurs at -10 MPa~\cite{Noblin2012}. 

Cavitation irrupts with a sudden bubble expansion. The inset in Fig. \ref{fig:  dyna} shows the growth of the cavitation bubble as a function of time, for several initial droplet sizes. The main panel in Fig. \ref{fig:  dyna} shows the result in dimensionless form, also including smaller bubbles. For all sizes we observe a power law for the initial growth stage, followed by (damped) oscillations and eventually a nearly static shape at $R/R_0 \approx 0.9$. A few minutes after the cavitation event, the bubble will eventually grow to $R/R_0 \approx 0.98$, practically recovering the initial droplet size. In Fig. \ref{fig:  dyna}, time is rescaled using $R_0/\sqrt{\mu /\rho}$, where $\sqrt{\mu /\rho}$ is the shear wave velocity. This appears to be the appropriate timescale for the cavity growth, showing that the dynamics emerges from the gel's inertia and elasticity. This is corroborated by the appearance of oscillations. 
However, they do not appear for all droplet/bubble sizes: the oscillations are not observed for the smallest droplet size ($R_0~=~$\SI{172}{~\micro\meter}), suggesting a transition from overdamped to underdamped oscillations.

The initial growth of all bubbles follow the same power law (Fig. \ref{fig:  dyna}). This can be explained using the expression for the kinetic energy, which for an incompressible medium surrounding a spherical bubble reads $\epsilon_{kin} = 2\pi\rho R^3 \dot{R}^2$. The term $\rho R^3$ represents the added mass of the cavity, and the same scaling should apply for bubbles that are not perfectly spherical. As cavitation occurs, the elastic energy is quickly released and we can assume the cavity grows at constant kinetic energy. This implies that $ \dot{R}^2 \sim 1/R^3$, and thus $R\sim t^{2/5}$. Indeed, this is the scaling observed experimentally for all bubbles (Fig. \ref{fig:  dyna}).  
From the prefactor of the growth law for the bubble radius, we can extract a typical value for the kinetic energy $\epsilon_{kin} \sim 0.5$~mJ. This can be compared to the elastic energy accumulated before cavitation, 3 mJ in the spherical approximation, which is indeed of the same order of magnitude. The difference between these values can be attributed to creasing (the spherical approximation is an overestimate), heat, and the vaporisation of the droplet.

\begin{figure}
  \includegraphics[width=0.48\textwidth]{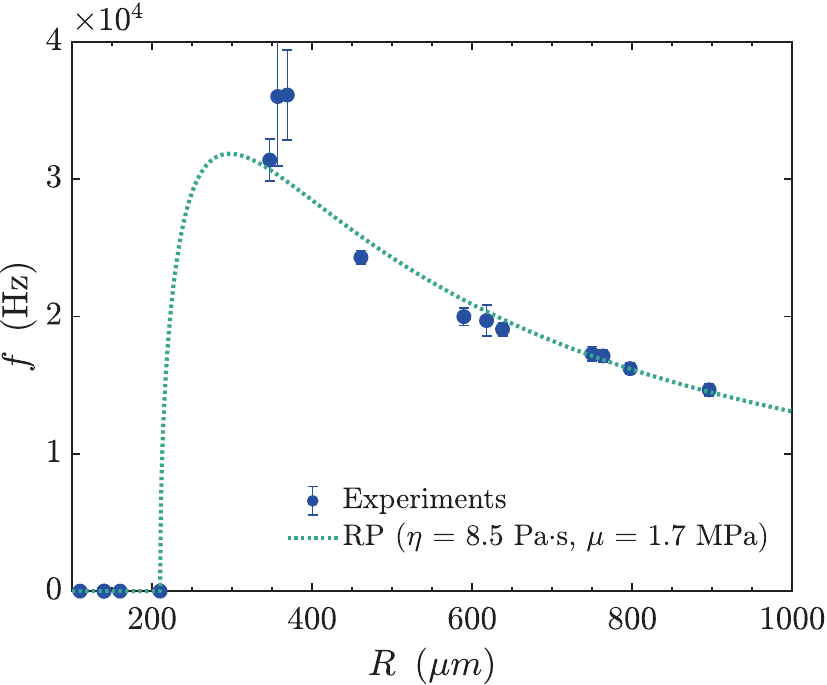}
\vspace{-3mm} \caption{Oscillation frequency of the cavitation bubble. The experimental data is shown together with the solution of the Rayleigh-Plesset equation (RP) including elasticity and viscosity (Eq. \ref{eq:omega}). The fitted values for the material parameters, $\mu$ and $\eta$, are also shown.   \label{fig:  osci} }
\end{figure}

The scaling of $R\sim t^{2/5}$ is the well-known inertial solution of the Rayleigh-Plesset equation for bubble dynamics in fluids. The modified Rayleigh-Plesset equation for bubbles in an elastic medium is given by 

\begin{equation}\label{eq:RP_full}
  \rho\left(R \ddot R+\frac{3}{2}\dot R^2\right)  = p_{cav}-p_\infty - \frac{2\gamma}{R}
  -\frac{\mu}{{\xi}^2}f'(\xi)-\frac{4\eta}{R}\dot{R},
\end{equation}
where we introduced an effective viscosity $\eta$ of the surrounding medium~\cite{Alekseev1999, Brennen1995}. The use of a damping of this ``viscous" form can be derived only for small deformations, where the effective viscosity 
can be inferred from the loss modulus as $G'' = \eta \omega$, see Supplemental Material. Similar formulations of the Rayleigh-Plesset equation in elastic media have been previously used in the context of forced bubble oscillations \cite{Gaudron2015, Estrada2018, Garbin2017,dollet2019bubble}. Considering the static solution $\dot R=0$ to this equation, we recover (\ref{p_bubble}) for the pressure inside the cavity. The oscillations observed in Fig. \ref{fig:  dyna} can now be analysed by considering small perturbations $R(t)=R_0(1+\epsilon e^{i\omega t})$, and linearising (\ref{eq:RP_full}). This gives an expression for the oscillation frequency,

\begin{equation}\label{eq:omega}
f \equiv \frac{\textrm{Re}(\omega)}{2\pi}  \simeq \frac{1}{\pi}\sqrt{\frac{\mu}{\rho R_0^2}-\frac{\eta^2}{\rho^2 R_0^4}},
\end{equation}
accounting for elasticity and viscous damping. Here we omitted the subdominant contributions due to pressure and surface tension (respectively smaller by $10^{-1}$ and $10^{-3}$). The result (\ref{eq:omega}) indeed predicts a transition from overdamped to underdamped oscillations, occurring at a critical drop radius $R^*=\eta/\sqrt{\mu\rho}$. 
In Fig. \ref{fig:  osci} we present our experimental measurements for the frequency, obtained by fitting $R(t)$ to exponentially damped oscillations. Given that the bubble has not yet fully recovered, we defined the bubble radius as the mean of the maximum and minimum value of the first oscillation. To compare the experimental data to (\ref{eq:omega}), we impose the experimental value of the critical drop radius, $R^*=$\SI{210}{\micro\meter}, and subsequently use the (dynamic) shear modulus $\mu$ as adjustable parameter. The values found ($\mu = 1.7$ MPa and $\eta = $ 8.5 Pa$\cdot$s) match very well with available literature on rheological measurements of PDMS at high frequencies \cite{Placet2015} and with our own measurements (see Supplemental Material). The excellent agreement in Fig. \ref{fig:  osci} shows that the oscillations are quantitatively captured by a viscoelastic Rayleigh-Plesset-type model.

In summary, we have studied a cavitation process similar to that found in the cells of the fern sporangia. Our experiments capture and quantify the slow diffusive evaporation, which leads to a build-up of negative elastic pressure that eventually leads to a very fast cavitation. We provided detailed insight into the cavitation dynamics, which was accurately described by an inertial-(visco)elastic model. 
Understanding the mechanism by which certain plants are able to transform enthalpy into mechanical energy could lead a way to design synthetic materials able to perform similar fast motions making use of the this smart energy transformation found in nature.



\emph{Acknowledgements - } The authors acknowledge fruitful discussions with Benjamin Dollet, Sascha Hilgenfeldt, Javier Rodr\'iguez-Rodr\'iguez and Jens Eggers. M.B. and A.M. acknowledge financial support from ERC (the European Research Council) Starting Grant No. 678573 \emph{NanoPacks}. M.C. and J.H.S. acknowledge financial support from ERC Consolidator Grant No. 616918 \emph{Soft Wetting}.

\bibliographystyle{apsrev4-1}
\bibliography{BibCavitation3}

\end{document}


\preprint{APS/123-QED}

\title{SUPPLEMENTARY MATERIAL of:\\
''Turning drops into bubbles: Elastic cavitation by diffusion''}

\author{M.A.Bruning}
\affiliation{%
 Physics of Fluids Group, Faculty of Science and Technology, Mesa+ Institute, University of Twente, 7500 AE Enschede, The Netherlands
}%
 \author{M. Costalonga}
\affiliation{Department of Mechanical Engineering, Massachusetts Institute of Technology, Cambridge, Massachusetts, USA}
\author{J. H. Snoeijer}%
\author{A. Marin}
 
\affiliation{%
 Physics of Fluids Group, Faculty of Science and Technology, Mesa+ Institute, University of Twente, 7500 AE Enschede, The Netherlands
}%

\date{\today}

\begin{abstract}

Rheological measurements of the PDMS-gel are performed on an Anton Paar MCR 502 rheometer, in the parallel plate configuration (PP25) and gap width 0.5 mm. The gel is cured in the rheometer at 70$^\circ$C for 100 minutes, while the storage modulus ($G'$) and loss modulus ($G''$) are monitored. 

After curing the sample, the two moduli are measured for a wide range of $\omega$: $10^{-2}-10^{2}$ rad/s, at a deformation of 0.5\%. These measurements are performed at three different temperatures: -15, 20 and 70$^\circ$C. Using the principle of time$-$temperature superposition, we are able to create one master curve at 70$^\circ$C, extending the upper limit of $\omega$ above $10^{5}$ rad/s. The results for $G'$ and $G''$ as a function of $\omega$ after this superposition are shown in Fig. 1.

In the measured range of $\omega$, the storage modulus remains almost constant, whereas the loss modulus changes significantly. From the loss modulus we can calculate the viscosity, using $G'' = \eta \omega$, which is frequency dependent. The range of frequencies of the cavitation bubbles is 10 to 40 kHz (see Fig. 4 in the main Letter), which is similar to a spread in $\eta$ of 4 Pa$\cdot$s. As the experimental frequency range is small, we assume a single value for $\eta$ to describe the experimental measurements. Note that beyond $R^*$, the exact value of $\eta$ has a relatively weak effect on the prediction (3).



\end{abstract}

\maketitle


 \begin{figure}[t]
  \includegraphics[width=0.9\textwidth]{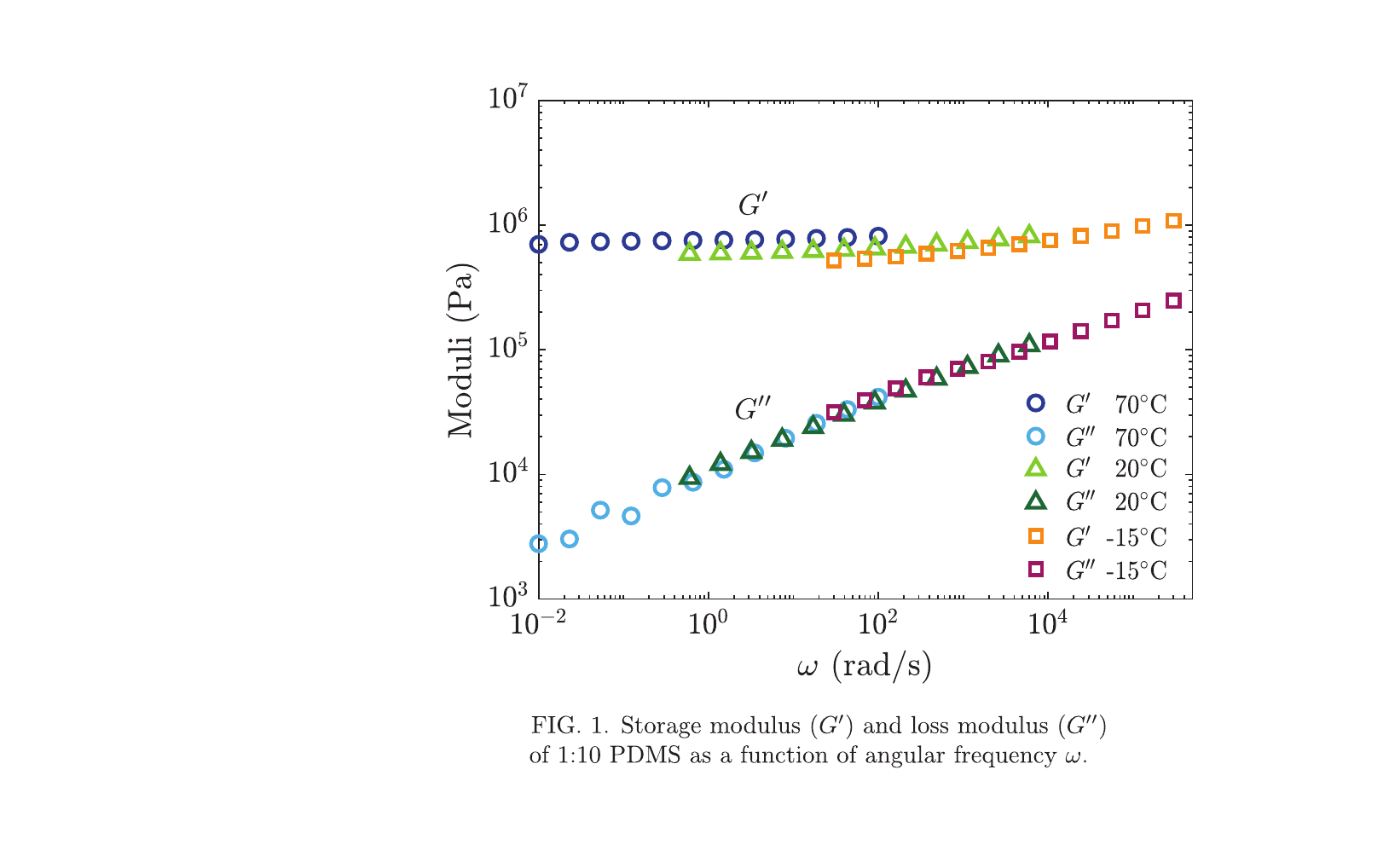}
\vspace{-3mm}
 
\end{figure}